\documentclass[11pt]{article}
\usepackage{epsfig}
\usepackage{amsmath}
\usepackage{amsfonts}

\begin{document}

\begin{titlepage}
 \begin{normalsize}
  \begin{flushright}
   {\tt hep-th/0408014}\\
   UT-04-21\\
   August 2004
  \end{flushright}
 \end{normalsize}
 \begin{Large}
  \vspace{0.6cm}
  \begin{center}
   \bf
   Semiclassical Strings on $AdS_5\times S^5/{\mathbb Z}_M$ and\\
   Operators in Orbifold Field Theories
  \end{center}
 \end{Large}
 \vspace{15mm}
 
 \begin{center}
  {\large
  Kota Ideguchi
  \footnote{E-mail :
  ideguchi@hep-th.phys.s.u-tokyo.ac.jp}
  
  \vspace{16mm}
  
  {\it Department of Physics,}\\[0.7ex]
  {\it University of Tokyo,}\\[0.7ex]
  {\it Hongo, Tokyo 113-0033, Japan}
  }
  \vspace{1.5cm}
 \end{center}
 \begin{abstract}
  We show agreements, at one-loop level of field theory, between
  energies of semiclassical string states on
  $AdS_5\times S^5/{\mathbb Z}_M$ and anomalous dimensions of operators
  in ${\cal N}=0,1,2$ orbifold field theories originating from ${\cal
  N}=4$ SYM. On field theory side, one-loop anomalous dimension matrices can be
  regarded as Hamiltonians of spin chains with twisted boundary
  conditions. These are solvable by Bethe ansatz. On string side,
  twisted sectors emerge and we obtain some string configurations in
  twisted sectors. In $SU(2)$
  subsectors, we compare anomalous
  dimensions with string energies and see
  agreements. We also see agreements between sigma models of both sides
  in $SU(2)$ and $SU(3)$ subsectors. 
 \end{abstract}

\end{titlepage}

\section{Introduction}
Recently, in context of AdS/CFT, energies of semiclassical string states
on $AdS_5\times S^5$ have successfully compared with anomalous dimensions
of operators in planar ${\cal N}=4$ Super Yang-Mills theory. 
In \cite{Gubser:2002tv}, it was shown 
that energy of semiclassical string states with large angular momentum
in $AdS_5$ agrees qualitatively with anomalous dimension of twist two
operators of ${\cal N}=4$ SYM. On the string side, the analysis of semiclassical
string states was
generalized to cases of strings having angular momentum in $S^5$ in
\cite{Frolov:2002av, Tseytlin:2002ny}. 
Then, some kind of semiclassical string states
with several large angular momentums in $S^5$ (and $AdS_5$) have been
found to have energies which have analytic expansion in
$\lambda/J^2$ ($\lambda$ : `t Hooft coupling, $J$: angular
momentum)\cite{Frolov:2003qc,Frolov:2003tu,Arutyunov:2003uj,Arutyunov:2003za,Tseytlin:2003ii}. 

On the other hand, it was shown that one-loop anomalous dimension matrix of
operators consisting only of scalars in ${\cal N}=4$ SYM is identical to
a Hamiltonian of $SO(6)$ spin chain which is solvable by Bethe
ansatz\cite{Minahan:2002ve}. This was generalized to $SU(2,2|4)$ super
spin chain containing
full sector of operators\cite{Beisert:2003yb, Beisert:2003jj}.
Interestingly, integrability is likely to hold at higher
loops\cite{Beisert:2003tq, Beisert:2003jb, Beisert:2003ys}. Bethe ansatz
including higher loops were investigated in \cite{Serban:2004jf,
Beisert:2004hm, Arutyunov:2004vx}

The Bethe equations of spin chain are difficult to solve in general. In
large number of sites limit, these equations are transformed to
integral equations
and can be solved. Then, the anomalous dimensions of the CFT operators can be
obtained and these agree quantitatively with the energies of the semiclassical
strings in $SU(2)$ subsector at one-loop level\cite{Beisert:2003xu,
Frolov:2003xy, Beisert:2003ea}. Agreement at higher loops was
investigated in \cite{Serban:2004jf, Beisert:2004hm}. This agreement holds at
2-loop level but there is discrepancy between string and gauge sides at
3-loop level. Some discussions of the origin of this phenomena were done
in these papers.

Moreover, an agreement between string and gauge sides
was confirmed in the sigma model level\cite{Kruczenski:2003gt,
Kruczenski:2004kw, Dimov:2004qv, Hernandez:2004uw, Stefanski:2004cw,
Kruczenski:2004cn} and general
solution level\cite{Kazakov:2004qf}. Other interesting
works concerning with semiclassical strings, anomalous dimensions and
spin chains were done in many 
papers\cite{Arutyunov:2003rg}-\cite{Mandal:2002fs}.

It is important to generalize the arguments of these agreements or
integrabilities to other models. There are several simple ways of
constructing these
models; orbifolding\cite{Wang:2003cu} or deforming\cite{Roiban:2003dw,
Berenstein:2004ys} the ${\cal N}=4$ case, adding
D-branes(open
strings)\cite{Stefanski:2003qr,Chen:2004mu,DeWolfe:2004zt,Chen:2004yf},
etc.(For other models see also \cite{Belitsky:2003ys, Belitsky:2004cz,
Alishahiha:2002fi, Bigazzi:2004yt, Ferretti:2004ba, DiVecchia:2004jw}.) 
In \cite{Wang:2003cu}, integrabilities of 
orbifold field theories and their deformations from conformal fixed
point were investigated purely in field theory point of view.

In this paper, we consider models obtained by orbifolding the ${\cal
N}=4$ case. In untwisted sector, a story is
almost the same as the original
case. The matter in orbifold cases is whether boundary condition of
twisted sector is compatible with integrability or not. If boundary
condition does not break integrability, it is expected that, using this
integrability, we can
confirm agreements of string and gauge sides even for twisted sector.
Indeed, this is the case and what we will do in this paper. 

We analyze ${\mathbb C}^3/{\mathbb Z}_M$ orbifold cases at one-loop
level of field theory. Our
analysis is applied to ${\cal N} = 0,1,2$ cases. 
We show integrabilities of one-loop anomalous dimension
matrix in these models. Interestingly, we find that the integrabilities
hold even in
sectors corresponding to broken global symmetries. These sectors are
described by solvable spin chains with twisted boundary conditions. 
Then, solving Bethe equations, we show agreements between energies of
string states and anomalous dimensions of CFT operators in $SU(2)$ subsectors. 
We also show
agreements between sigma models of string and gauge sides in $SU(2)$ and 
$SU(3)$ subsectors. 

Our paper is organized as follows. In section 2, we consider $SU(2)$
subsectors of ${\cal N}=2$ cases. On string side, we construct 2-spin
circular strings on $AdS_5\times S^5/{\mathbb Z}_M$. On gauge side, we
see integrabilities of one-loop anomalous dimension matrices
corresponding to $SU(2)$ sectors. This integrabilities hold for
sectors corresponding to broken symmetries, as well as for a sector
corresponding to a global symmetry. Then, we show agreements between string
energies and anomalous dimensions for rational cases. In section 3, we
generalize this analysis to $SU(2)$ subsectors of more general
orbifolds including ${\cal N}=0,1$ cases. Then, an
agreement between sigma models of both sides is
confirmed. In section 4, we extend this agreement of sigma models to
$SU(3)$ subsectors. 
In section 5, we conclude this paper.

\section{$SU(2)$ sectors: ${\cal N}=2$ cases}

We consider ${\mathbb C} \times {\mathbb C}^2/{\mathbb Z}_M$
orbifolds in this section. The coordinate of ${\mathbb C}^3$ is $(X,\ Y,\ Z)$. There is $SO(6)$ rotation symmetry on ${\mathbb C}^3$. This
group have subgroup $SU(2)_L\times SU(2)_R\times U(1)_R$. $U(1)_R$
is the rotation in the $Z$-plane. 
$(X,\ Y)$ forms a doublet of $SU(2)_L$ and $(X,\
Y^\dagger)$ forms a doublet of $SU(2)_R$.
We take ${\mathbb Z}_M \subset
SU(2)_L$ and this choice breaks supersymmetry to ${\cal
N}=2$. Our orbifold is described as ${\mathbb C}^3$ with the
following identification,
\begin{equation}
( X,\ Y,\ Z) \sim ( w X,\ w^{-1}Y,\ Z), \quad w= e^{2\pi i/M}. \label{orb}
\end{equation}
The above orbifolding breaks
the rotation symmetry to $U(1)_L\times SU(2)_R\times U(1)_R$. 

The near horizon geometry of the D3-branes located on the fixed point of
the orbifold is $AdS_5\times S^5/{\mathbb Z}_M$\cite{Kachru:1998ys}. The CFT dual of this
background is a ${\cal N}=2$ quiver gauge theory.

\subsection{Semiclassical string rotating in $S^5/{\mathbb Z}_M$}
We consider semiclassical string configuration in $AdS_5 \times
S^5/{\mathbb Z}_M$. Especially, we are interested in configurations of twisted
sectors. In fact, there are a simple generalization of
semiclassical configuration in \cite{Arutyunov:2003za} to solutions in
the twisted sectors. In \cite{Arutyunov:2003za}, they treat the $O(6)\times SO(4,2)$
sigma model as a bosonic part of strings propagating in
$AdS_5\times S^5$. When considering classical motion of the strings
which is nontrivial only in $S^5$, the action is reduced to time
direction in $AdS_5$ and the $O(6)$ sigma model. 
\begin{equation}
 S = \frac{R^2}{2\pi\alpha'} \int d\tau d\sigma 
\left( \frac{1}{2}(\partial_a t)(\partial^a t) - 
\frac{1}{2} (\partial_a X_\mu) (\partial^a X_\mu)
+ \frac{1}{2} \Lambda (X_\mu X_\mu -1)\right) ,
\end{equation}
where $\mu= 1,\dots, 6$ and 
\begin{equation}
R^4 = 4\pi g_s NM {\alpha'}^2, \quad g_{QGT}^2 = 4\pi g_s M.
\end{equation}
Hence, the overall factor of the action is $\frac{R^2}{2\pi \alpha'} =
\frac{\sqrt{g_{QGT}^2 N}}{{2\pi}} \equiv
\frac{\sqrt{\lambda}}{{2\pi}}$. 
Equations of motion are 
\begin{equation}
 \partial^2 t=0,\quad 
 \partial^2 X_\mu + \Lambda X_\mu=0,\quad 
  X_\mu X_\mu =1.
\end{equation}
One can take the following ansatz, 
\begin{gather}
 t = \kappa \tau, 
 \quad X \equiv X_1 +i X_2= r_1(\sigma) e^{i (\alpha_1(\sigma) + w_1 \tau)}, \\
 Y \equiv X_3+ iX_4 = r_2(\sigma) e^{i (\alpha_2(\sigma) + w_2 \tau)},
 \quad Z \equiv X_5+ iX_6= r_3(\sigma) e^{i (\alpha_3(\sigma) + w_3 \tau)}.
\end{gather}
$r_i,\; \alpha_i$ satisfy the following periodicity conditions. 
\begin{equation}
 r_i(\sigma+ 2\pi) = r_i(\sigma), \quad \alpha_i(\sigma+2\pi) = 
  \alpha_i(\sigma) + 2\pi m_i,
\end{equation}
where
\begin{equation}
 m_1 = \frac{k}{M} + \tilde{m}_1, \quad m_2 = -\frac{k}{M} + \tilde{m}_2, 
  \quad m_3 = \tilde{m}_3, \quad \tilde{m}_i \in Z,\ k= 0,1,\dots ,M-1.
\end{equation}
In $S^5$ case, $m_i$ takes integer value, but in orbifold case we can
take fractional values corresponding to twisted sectors. With
this difference, remaining arguments is the same as that of $S^5$
case. 

The most simple examples are circular strings. We consider these
configurations now. These solutions
take the following forms,
\begin{equation}
 r_i(\sigma) = a_i= \text{const}, \quad \alpha_i = m_i \sigma, 
\quad \Lambda = -\nu^2. 
\end{equation}
The equations of motion impose the relations between parameters,
\begin{equation}
 \omega_i^2 = m_i^2 + \nu^2, 
  \quad \sum_{i=1}^3 a_i^2 = 1. \label{Lag}
\end{equation}
Virasoro constraints become 
\begin{equation}
 \kappa^2 = \sum_{i=1}^3 a_i^2 ( \omega_i^2 + m_i^2),\quad 
  \sum_{i=1}^3 a_i^2 \omega_i m_i=0. \label{Vira}
\end{equation}
A string energy $E$ and angular momentums $J_i$ are expressed in terms
of sigma model variables,
\begin{gather}
 {\cal E} \equiv \frac{E}{\sqrt{\lambda}}= \int \frac{d\sigma}{2\pi}
 \dot t = \kappa, \nonumber\\
 {\cal J}_i \equiv \frac{J_i}{\sqrt{\lambda}} = \int \frac{d\sigma}{2\pi} 
(X_{2i -1} \dot X_{2i} - X_{2i} \dot X_{2i-1}) =a_i^2 \omega_i. 
\quad (i = 1,2,3) \label{ESsigma}
\end{gather}
$J_1$, $J_2$ and $J_3$ are the rotations in $X,\ Y,\ Z$-planes
respectively. 
We define a total spin $L = \sum_{i=1}^3 |J_i|$. 

The relations \eqref{Vira} and \eqref{Lag} can be rewritten in terms
of the energy and the spins \eqref{ESsigma}, 
\begin{equation}
{\cal E}^2 = 2\sum_{i=1}^3 \sqrt{m_i^2 +\nu^2}|{\cal J}_i| -\nu^2, 
\quad \sum_{i=1}^3 \frac{|{\cal J}_i|}{\sqrt{m_i^2 + \nu^2}}=1, 
\quad \sum_{i=1}^3 m_i {\cal J}_i =0. \label{eq14}
\end{equation}
In order to describe ${\cal E}$ as a function of $m_i$ and ${\cal J}_i$,
firstly using the second equation we write $\nu$ as a function of $m_i$
and ${\cal J}_i$ in large ${\cal J} \equiv L/\sqrt{\lambda}$ expansion. Then,
inserting this $\nu$ to the first equation, we obtain ${\cal E}$ as a
function of $m_i,\ {\cal J}_i$. The third equation is imposed at final
stage. 

In orbifold theory, we have to restrict string states to invariant
states respect to orbifold action $e^{2\pi i (J_1 - J_2)}$. 
So, a projection condition is 
\begin{equation}
 J_1 - J_2 = 0 \mod M. \label{invcd}
\end{equation}

In order to compare string energies with anomalous dimensions of CFT
operators in $SU(2)$ subsector, we need 2-spin solutions of
strings. Here, we firstly consider string configurations extending in $X$ and
$Y$ directions and then consider configurations extending in $X$ and $Z$ directions.

\paragraph{${\cal J}_1, {\cal J}_2 \neq 0,\ {\cal J}_3 =0, a_3 =0$ case}
The string energy has the following expansion in large 
${\cal J}$, 
\begin{equation}
 {\cal E} = {\cal J} + \frac{|m_1||m_2|}{2{\cal J}} + \cdots.\label{eq15}
\end{equation}
The third equation of \eqref{eq14} becomes 
\begin{equation}
m_1 {\cal J}_1 + m_2 {\cal J}_2 =0.\label{eq16}
\end{equation}
When ${\cal J}_1 > 0 > {\cal J}_2$, $m_2$ has the same sign as $m_1$.
Therefore the parameters $\tilde{m}_i$ should satisfy 
the condition $(\tilde{m}_1\geq 0,\ \tilde{m}_2 > 0)$ or 
$(\tilde{m}_1 <0,\ \tilde{m}_2 \leq 0)$. The energy has the form, 
\begin{equation}
 E = L + \frac{\lambda}{2L}(\tilde{m}_1 + \frac{k}{M}) 
(\tilde{m}_2 - \frac{k}{M}). \label{energy1}
\end{equation}
In field theory side, by an argument of classical scaling dimension and
charges, the CFT operators dual to these
strings should have
$J_1$ $X$s and $|J_2|$ $Y^\dagger$s. 
This sector correspond to $SU(2)_R$ global symmetry of the orbifold
theory. We will see spin chain corresponding to this sector in sec.\ 2.2.2.

When ${\cal J}_1, {\cal J}_2>0$, then $m_2$ has the opposite sign to
$m_1$. In this case, $( \tilde{m}_1 \geq 0,\ \tilde{m}_2
\leq 0)$ or $(\tilde{m}_1 <0,\ \tilde{m}_2 >0)$ should be
satisfied. The string energy is 
\begin{equation}
 E = L + \frac{\lambda}{2L} (\tilde{m}_1 + \frac{k}{M})
(\hat{m}_2 + \frac{k}{M}), \label{energy2}
\end{equation}
where $\hat{m}_2 \equiv -\tilde{m}_2$, so that $\hat{m}_2$ has the same
sign as $\tilde{m}_1$. 
By an argument of classical dimension and charges, these string configurations should
correspond to $SU(2)_L$ sector consisting of fields $X$ and $Y$ of the field theory. The $SU(2)_L$ global
symmetry of the original theory is
broken in the orbifold theory. We will see spin chain of this sector in
sec 2.2.3.

\paragraph{${\cal J}_1, {\cal J}_3\neq 0$, ${\cal J}_2=0$, $a_2 =0$
    case} 
With replacing $m_2$, ${\cal J}_2$ by $m_3$, ${\cal J}_3$ respectively, the equations
\eqref{eq15}, \eqref{eq16} hold.
We assume
$J_1,J_3>0$. In this case, 
$m_1 = \tilde{m}_1 + \frac{k}{M}$, $m_3 = \tilde{m_3}$, $\tilde{m_i} \in
Z$, and $(\tilde m_1 \geq 0,\ \tilde m_3 <0)$ or $(\tilde m_1 <0,\ \tilde m_3
>0)$ should be satisfied. The energy becomes
\begin{equation}
 E = L + \frac{\lambda}{2L}(\tilde{m}_1 + \frac{k}{M})\hat{m}_3, 
  \label{energy3}
\end{equation}
where $\hat{m}_3 \equiv -\tilde{m}_3$, so that $\hat{m}_3$ has the same
sign as $\tilde{m}_1$. These strings should correspond to the operators
consisting of $X$ and $Z$. We will see spin chain of this sector in the
last paragraph of sec 2.2.3.

\subsection{Gauge theory side}

\subsubsection{General aspects}
In this section, we discuss general aspects of the gauge side, not
restricted to $SU(2)$ sectors. 
We consider orbifold field theories originating from ${\cal N}=4$ super Yang-Mills theory. 
Starting from $U(NM)$ ${\cal N}=4$ SYM, we project
fields in ${\cal N}=4$ theory onto $Z_M$ invariant components. 
Fields surviving the projection satisfy 
\begin{equation}
\phi= \gamma^\dagger (r\cdot \phi )\gamma,
\end{equation}
where $\gamma = \text{diag} (1,\omega, \dots, \omega^{M-1})$ acting on
gauge index and $r$ is representation of orbifold
action on fields with internal space indices. 
At the beginning, all fields in ${\cal N}=4$ SYM are adjoint
representation and $NM \times NM$ matrices. 
Due to this projection, the gauge groups become
$U(N)_1\times U(N)_2 \times \cdots \times U(N)_M$. 
The $Z_M$
actions on the ${\mathbb C}^3$ for ${\cal N}=2$ case are indicated in
\eqref{orb}. So, the projection conditions for the complex scalar fields are 
\begin{equation}
 \gamma^\dagger X \gamma = \omega X,\ \gamma^\dagger Y \gamma 
= \omega^{-1} Y,\ \gamma^\dagger Z \gamma = Z
\end{equation}
The components which survive the projection are the following.
\begin{gather}
Z = \begin{pmatrix}
Z_{11} & & & \\
& Z_{22} & & \\
& & \ddots& \\
& & & Z_{MM} 
\end{pmatrix}, 
X = \begin{pmatrix}
0& X_{12}& & \\
& 0& X_{23}& \\
& & \ddots& \ddots\\
X_{M1}& & &
\end{pmatrix},
Y = \begin{pmatrix}
0& & &Y_{1M}\\
Y_{21}& 0& &\\
&  Y_{32} &0 &\\
& & \ddots&\ddots
\end{pmatrix}
\label{scalar}
\end{gather}
Each component is $N\times N$ matrix and is complex scalar part of chiral
superfield in terms of ${\cal N}=1$ supersymmetry.
$Z_{i,i}$ is adjoint scalar field of the $i$-th U(N) gauge group and scalar
component of vector multiplet in ${\cal N}=2$ context. 
$X_{i,(i+1)}$ is bifundamental representation of gauge groups, $N$ for $i$-th
gauge group and $\bar{N}$ for $i+1$-th gauge group\footnote{We denote
this type of bifundamental representation as $(i,
\overline{i+1})$.}. $Y_{(i+1),i}$ is bifundamental representation, $\bar{N}$
for $i$-th gauge group and $N$ for $i+1$-th gauge group. The two chiral
superfields, corresponding to $X_{i,(i+1)}$ and $Y_{(i+1), i}$, form
hypermultiplet in bifundamental representation in ${\cal N}=2$ context. 
$X_{i,(i+1)}$ and the Hermite conjugate of $Y_{(i+1),i}$ form a doublet of
$SU(2)_R$ symmetry. This $R$-symmetry is a part of $SO(6)$ R-symmetry
of the original theory and remains as a global symmetry in the orbifold field
theory. On the other hand, $X$ and $Y$ have formed a doublet of
$SU(2)_L\subset SO(6)$ in the ${\cal N}=4$ theory. However, this $SU(2)_L$
symmetry is broken by orbifolding and is not global symmetry of the orbifold
theory. \\

Arguments of the rest of this subsection is applied not only to the ${\cal N}=2$ cases but also to
more general orbifold cases in Section 3. 

In our paper, We does not decompose the $NM\times NM$ matrix to $N\times
N$ matrix. That is, we treat the single trace operators which have the
following form, 
\begin{equation}
 \text{tr} XXYZY^{\dagger}X \cdots,\label{eq23}
\end{equation}
where $X$, $Y$ and $Z$ are $NM\times NM$ matrices of
\eqref{scalar} and the trace is gauge index trace. Using this operator, we
need not to specify which gauge group each field belongs to. More
carefully, in order to take account of all gauge invariant
operators we have to introduce the `twisted sectors' of the
operators,
\begin{equation}
 \text{tr} \gamma_k XXYZY^{\dagger}X \cdots,
\end{equation}
where 
\begin{equation}
 \gamma_k = \text{diag} (1, w^k, w^{2k}, \dots, w^{(M-1)k}), \quad w=
  e^{2\pi i/M}. \quad k = 1, 2, \dots, M-1
\end{equation}
We call these operators twisted operators. On the other hand, the
operators of \eqref{eq23} correspond to $k=0$ and we call these
untwisted operators. 

The operators with different twist number $k$ are
not mixed in 2-point function. This is due to ${\mathbb Z}_M$ symmetry which orbifold
field theory has. An element $g=e^{2\pi i/M}$ in this ${\mathbb Z}_M$ acts on
arbitrary field $\phi_{i,j}$ of field theory as $g: \phi_{i,j} \rightarrow
\phi_{i+1, j+1}$.($i$ and $j$ denote $i$-th gauge group and $j$-th gauge
group, respectively.) Then, this acts on single trace operator as $g:
\text{tr} \gamma_k XXY\cdots \rightarrow \omega^{-k} \text{tr} \gamma_k
XXY\cdots$. This means that operators with twist number $k$ are
eigenstate of this ${\mathbb Z}_M$ symmetry and have eigenvalue $\omega^{-k}$ for the
element $g$. Then, 
2-point function between operators with different twist number, vanishes. Hence, we can treat each sector separately. 

The trace of operators requires that $SO(6)$ charge $(J_1,\ J_2,\ J_3)$ of
operators should satisfy $J_1 - J_2 = 0 \mod M$ in the ${\cal N}=2$ case. If
otherwise, diagonal parts of the operators are zero matrices and the
operators vanish by the trace. This fact can be seen as following. 
\begin{align}
 \text{tr}\gamma_k XYXY^\dagger Z\cdots =& 
 \text{tr}\gamma_k \gamma_1^{-1} \gamma_1 XYXY^\dagger
 Z\cdots = \omega^{-1} \text{tr} \gamma_k \gamma_1^{-1} 
 X\gamma_1 YXY^\dagger Z \cdots \nonumber\\
 =& \omega^{-(1-1)}
 \text{tr} \gamma_k\gamma_1^{-1} XY\gamma_1
 XY^\dagger Z \cdots 
 = \cdots =
 \omega^{-(J_1-J_2)}\text{tr} \gamma_1\gamma_k\gamma_1^{-1}
 XYXY^\dagger Z\cdots \nonumber\\
 =& \omega^{-(J_1-J_2)} \text{tr}\gamma_k XYXY^\dagger Z\cdots. 
\end{align}
So, the invariant condition \eqref{invcd}
is automatically satisfied in field theory side. It is easy to
generalize this fact to more general ${\mathbb C}^3 /{\mathbb Z}_M$ orbifold
cases in Section 3. 

In the planar limit, 
anomalous dimension matrix of untwisted operators ($k=0$) in orbifold theory is
the same as that of ${\cal N}=4$ theory. 
In orbifold theory, vertices of Feynman diagram are
inherited from ${\cal N}=4$ SYM. Hence, the number of fields of each Lorentz
type which run loops are the same as that of diagrams in
${\cal N}=4\ U(N)$ SYM. Then, we obtain the same anomalous
dimension matrix as ${\cal N}=4$ $SU(N)$ SYM case. We explain this fact in Appendix. 
This agreement could also be
explained by `inherited principle' of \cite{Bershadsky:1998cb,
Bershadsky:1998mb}. 

However, for twisted sectors of operators ($k\neq 0$) there is a subtlety about
interactions between the $L$-th site and
the first ($(L+1)$-th) site. That is, the interactions involving the $L$-th site and first site
step the matrix $\gamma_k$ and may generate additional phase
shift. Hence, the Feynman diagrams which have an interaction between
the $L$-th site and the first site may have these phase shifts. Whether this shift occurs or not depends on an order of gauge
group at the interaction point. When the phase shifts occur in some of
these diagrams, the anomalous dimension matrix is not
the same as that of untwisted sector. Nevertheless, we can resolve this
subtlety and we find that the spin chain systems corresponding to these
twisted sectors of operators are also solvable. We will see concrete examples of this
fact in the following subsections.

\subsubsection{XXX spin chain corresponding to $SU(2)_R$ symmetry}
Let us consider $SU(2)$ subsectors. $SU(2)$ subsectors, not
only $SU(2)_R$ but also others, are closed for loop corrections, the
feature inherited from the ${\cal N} =4$ SYM\cite{Beisert:2003tq}. 
We consider operators of $SU(2)_R$ subsector in this subsection. This
$SU(2)_R$ is a global symmetry of the orbifold theory we are
considering. The single trace operators in $SU(2)_R$ subsector have the
following form 
\begin{equation}
 \text{Tr} \gamma_k X^{J_1} {Y^\dagger}^{|J_2|} + permutations. 
  \quad (J_1>0, \ J_2 <0) \label{XYdag}
\end{equation}
$X$ and $Y^\dagger$ form a doublet of
$SU(2)_R$. These operators have the $SO(6)$ charges $(J_1,\ J_2,\ 0)$ 
and should correspond to the strings with these charges. The circular
strings with these charges are constructed in the above and have the
energy \eqref{energy1}. 

\begin{figure}[tb]
\begin{center}
\epsfig{file=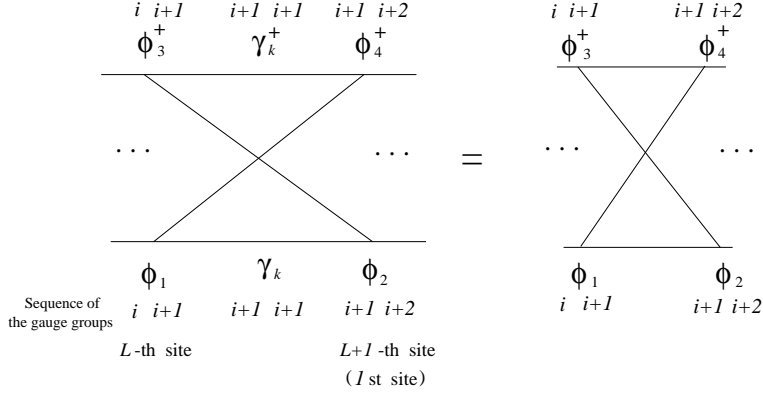}
\end{center}
\caption{Interaction between the $L$-th site and the first ($(L+1)$-th) site. Each
 $\phi$ is $X$ or $Y^{\dagger}$ and $i$, $i+1$ and $i+2$ represent the
 gauge groups.} 
\label{nonphase}
\end{figure}

We treat $X$ as an up-spin
and $Y^\dagger$ as a down-spin. More precisely, for the
first site, 
$\uparrow_{\text{1st site}}$($\downarrow_{\text{1st site}}$) is identified
as $\gamma_k X$($\gamma_k Y^\dagger$). Periodic boundary condition $\uparrow_{\text{L+1th
site}}= \uparrow_{\text{1st site}}$, $\downarrow_{\text{L+1th
site}}=\downarrow_{\text{1st site}}$ is imposed. 
We denote $\text{tr} (\gamma_k XY^\dagger XXY^\dagger \cdots)$ as
$|\uparrow\downarrow\uparrow\uparrow\downarrow\cdots\rangle_k$, where a
suffix $k$
means that this state corresponds to a operator with twist number $k$. 
We assume $J_1>|J_2|>0$ and take $|\uparrow \uparrow \cdots
\uparrow\rangle_k$ as ground state. 
Both of $X,\ Y^\dagger$ are
$(i, \overline{i+1})$ type
bi-fundamentals. So, the sequence of gauge groups $i$, which denote 
$i$-th gauge group, in the trace is
one-way. The phase shift for twisted sector mentioned above does not happen in this
case (See Fig.\ \ref{nonphase}). Then a one-loop anomalous dimension
matrix for any twist number sector of operators is the same as that of $SU(2)$
subsector in ${\cal N}=4$ SYM. 
\begin{equation}
 D_{\text{1-loop}} = \frac{\lambda}{16\pi^2} 
\sum_{l=1}^L (1-\vec\sigma_l \cdot \vec\sigma_{l+1}), \label{XXX}
\end{equation}
where $\vec{\sigma}_l$ are the Pauli matrices which act on the spin at
the $l$-th site. 
This takes the form of a Hamiltonian of the $\text{XXX}_{1/2}$ Heisenberg
spin chain. The boundary condition is periodic now. This model is
solvable by Bethe ansatz. Bethe equations are 
\begin{equation}
 (\frac{u_i + i/2}{u_i -i/2})^L = \prod_{j=1, j\neq i}^{|J_2|} \frac{u_i -
u_j + i}{u_i - u_j -i}. \label{bethe}
\end{equation}

Compared to ${\cal N}=4$ case, Only difference is
in a translation condition. In the presence of $\gamma_k$, the cyclicity of
the trace imposes the following condition,
\begin{equation}
\text{tr} \gamma_k \phi_1 \cdots \phi_L
=\text{tr}\phi_L \gamma_k \phi_1 \cdots 
= w^k \text{tr} \gamma_k \phi_L \phi_1 \cdots, 
\end{equation}
where $\phi_i$ represent $X$ or $Y^{\dagger}$ field. 
So, the translation condition for the spin chain states becomes 
\begin{equation}
T|\text{state}\rangle _k = w^{-k} |\text{state}\rangle _k,
\end{equation}
where $T=e^{iP}$ is the operator which translate each site to the right by a
site. In terms of Bethe roots this condition is
described as
\begin{equation}
 e^{iP}= \prod_{j=1}^{|J_2|} \frac{u_j + i/2}{u_j -i/2} 
= w^{-k}. \label{momentum}
\end{equation}

The anomalous dimensions, which are proportional to energies of
the spin chain, take the following form, 
\begin{equation}
 \gamma = \frac{\lambda}{8\pi^2} \sum_{i=1}^{|J_2|} \frac{1}{u_i^2 +1/4}. 
\label{anom}
\end{equation}
Rescaling $u_i = L x_i$ and taking large $L$ limit, the logarithms of
\eqref{bethe}, \eqref{momentum} become 
\begin{gather}
 \frac{1}{x_i} = \frac{2}{L} \sum_{j=1,j\neq i}^{|J_2|} 
\frac{1}{x_i - x_j} - 2\pi n.\quad n\in {\mathbb Z} \label{bethelog}\\
 \frac{1}{L} \sum_{i=1}^{|J_2|} \frac{1}{x_i} = -2\pi (m +
 \frac{k}{M}),\quad m \in {\mathbb Z}
\label{momentumlog}
\end{gather}
where we assume that all mode numbers of the Bethe equations are $n$. 
The anomalous dimension of \eqref{anom} becomes in the large $L$
expansion, 
\begin{gather}
\gamma = \frac{\lambda}{8\pi^2 L^2}\sum_{i=1}^{|J_2|} \frac{1}{x_i^2}.
\label{anomlargeL}
\end{gather}
The equations \eqref{bethelog}, \eqref{momentumlog} can be solved by
various methods\cite{Kazakov:2004qf, Lubcke:2004dg}. The
solutions are rational
type. We will outline one of the methods in Appendix. Using these solutions, we obtain the anomalous
dimensions from \eqref{anomlargeL}.(Also see Appendix.) Leading terms of these in large $L$
expansion are 
\begin{equation}
 \gamma = \frac{\lambda}{2L}(m+ k/M)(n-m - k/M).
\end{equation}
Note that $(m+k/M)$ has the same sign as $n$ and $(n-(m+k/M))$ also has
the same sign as $n$.(See Appendix.) 
Hence, identifying that $m_1 = m+ k/M, m_2 = n -(m+k/M)$ and remembering
that $m_2$ has the same sign as $m_1$, this result agrees
with \eqref{energy1}. Especially, we can see from this result that
the twisted sectors of the strings are dual to the twisted sectors of the
operators. 

\subsubsection{XXX spin chain with twisted boundary condition
corresponding to broken $SU(2)$ symmetry}
Next, we consider spin chains corresponding to $SU(2)$ symmetries
broken by orbifolding. 
One of examples is the $SU(2)_L$ sector, which consists of the
following operators, 
\begin{equation}
 \text{tr} \gamma_k X^{J_1} Y^{J_2} + permutations. 
\quad (J_1,\ J_2 >0)
\end{equation}
$X$ and $Y$ fields form a doublet of $SU(2)_L$ symmetry, which is broken
by orbifolding. 
These operators should correspond to string states with charges
$J_1,\ J_2>0$. Such a string state was constructed in the previous
section and the string energy was \eqref{energy2}. 
We identify X(Y) as up-spin(down-spin). As like in $SU(2)_R$ case, the
first site state
$\uparrow_{\text{1st site}}$($\downarrow_{\text{1st site}}$) is
identified $\gamma_k X$($\gamma_k Y$). A periodic boundary condition between
the first site and the $L+1$th site is imposed. We assume $J_1>J_2$ and
take $|\uparrow \uparrow \cdots \uparrow\rangle_k$ as the ground
state. 

\begin{figure}[tb]
\begin{center}
\epsfig{file=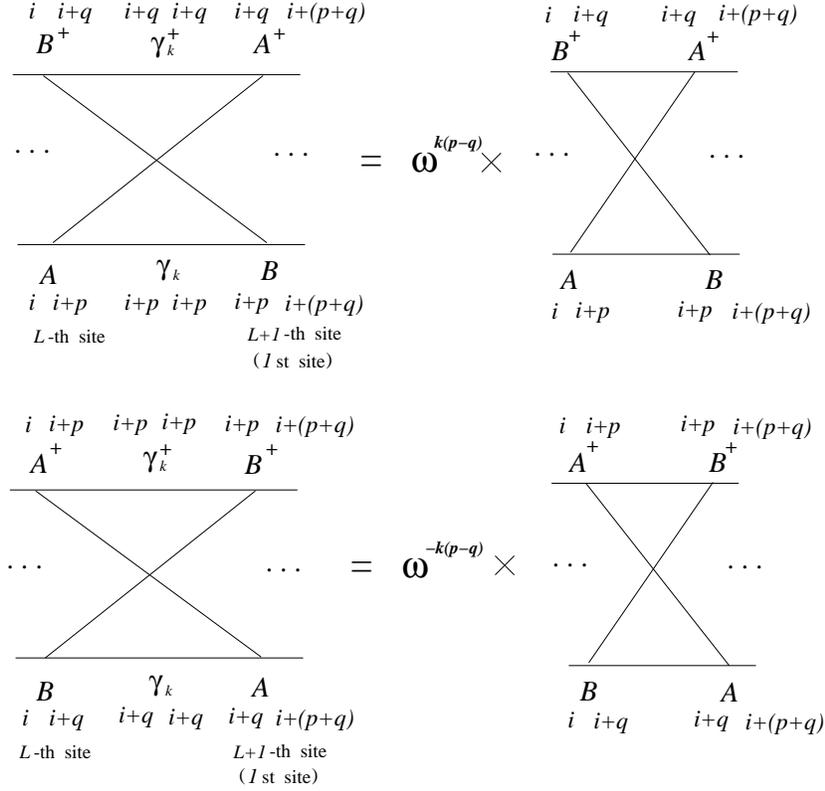}
\end{center}
\caption{The interactions which generate the phase shift at the end (the
 beginning) of the chain. $A$ and $B$ are fields consisting of
 operators. $A$ is $(i, \overline{i+p})$-type and $B$ is $(i,
 \overline{i+q})$-type operator. In the $SU(2)_L$ case, $A=X,\ B=Y$ and
 $p=1, q=-1$. }
\label{phase}
\end{figure}

$X$ field is a $(i,\ \overline{i+1})$-type operator but $Y$ field is a
$(i,\ \overline{i-1})$-type operator. Hence the sequence of the
gauge groups in the trace of the operator is not one-way but zigzag. This causes the
phase shifts of interactions between the $L$-th site and the first ($(L+1)$-th) site. Indeed the
interactions which interchange an up-spin and a down-spin generate the phase
shifts. At these interactions, the sequence of the gauge groups changes
between two traces of two operators which correlate with each other. This difference generates phase shift. These type of
interaction vertices are included by $F$-term potential
$\text{tr}|[X,Y]|^2$. The interactions which generate the phase shifts
are shown in Fig.\ \ref{phase}. On the other hand if the interaction is
proportional to the identity operator on the spin chain states, the sequence of the gauge groups does not
change between two traces and the phase shift does not occur.

In the existence of the phase shifts, the one-loop anomalous dimension
matrix does not agree with
XXX-spin chain Hamiltonian \eqref{XXX}. However, by changing the basis of
the spin chain to 
\begin{gather}
 |\tilde\downarrow_{\text{L+1-th site}}\rangle_k = u
  w^{-2k}|\downarrow_{\text{L+1-th site}}\rangle_k, 
  \quad |\tilde\uparrow_{\text{L+1-th site}}\rangle_k = 
  u|\uparrow_{\text{L+1-th site}}\rangle_k, \\
 |\tilde\downarrow_{\text{i-th site}}\rangle_k= 
  |\downarrow_{\text{i-th site}}\rangle_k, 
  \quad |\tilde\uparrow_{\text{i-th site}}\rangle_k = 
  |\uparrow_{\text{i-th site}}\rangle_k, \quad (i = 1, \dots, L)
\end{gather}
the phase shifts are cancelled and the one-loop anomalous dimension matrix becomes the same as
\eqref{XXX}. $u$ is an arbitrary
phase. In other words, we can regard our model as XXX spin chain of
$|\tilde{\uparrow}\rangle$ and $|\tilde{\downarrow}\rangle$ with
Hamiltonian \eqref{XXX} and with twisted boundary condition, 
\begin{equation}
|\tilde\downarrow_{\text{1st site}}\rangle_k = 
u^{-1} w^{2k} |\tilde\downarrow_{\text{L+1-th site}}\rangle_k, 
\quad |\tilde\uparrow_{\text{1st site}}\rangle_k = 
u^{-1} |\tilde\uparrow_{\text{L+1-th site}}\rangle_k. \label{spredef}
\end{equation}
The Hamiltonian preserves $SU(2)$ symmetry, but this boundary condition
breaks this symmetry. 
This system is exactly solvable by Beth
ansatz with a slight modification\cite{Berenstein:2004ys}. Since it is convenient for the ground
state to be periodic along the spin chain, we choose $u=1$. With the
twisted boundary condition, Bethe
equations become 
\begin{equation}
 w^{2k} \left(\frac{u_i + i/2}{u_i-i/2}\right)^L 
= \prod_{j=1,j\neq i}^{J_2} \frac{u_i- u_j +i}{u_i -u_j -i}.\label{bethe2}
\end{equation}
Only difference to the periodic case is phase factor $w^{2k}$ in the
l.h.s., which reflects twisted boundary condition \eqref{spredef}. 
The cyclicity of trace indicates
\begin{equation}
 \text{tr} \gamma_k XY \cdots \phi 
= w^{k\sigma_3^{\phi}}\text{tr} \gamma_k \phi XY \cdots,
\end{equation}
where $\phi$ is $X$ or $Y$ field and $\sigma_3^{X}=1$ and $\sigma_3^{Y} =
-1$. This relation can be rewritten in terms of the spin chain,
\begin{align}
 |\uparrow \downarrow \cdots \phi\rangle_k 
 &= w^{k\sigma_3^{\phi}}|\phi \uparrow \downarrow \cdots\rangle_k\nonumber\\
 &= w^{k\sigma_3^{\phi}} w^{k(1-\sigma_{3}^\phi)} 
 T |\uparrow \downarrow \cdots \phi\rangle_k 
 = w^k T| \uparrow\downarrow\cdots \phi\rangle_k,\label{trans2}
\end{align}
where $\phi$ represents $\uparrow$ or $\downarrow$ appropriately.
When going to the second line, we used the twisted boundary condition, 
$|\phi_{\text{1st site}}\rangle_k 
= w^{k(1-\sigma_3^{\phi})}|\phi_{\text{L+1-th site}}\rangle_k$.
Using the Bethe roots, the equation \eqref{trans2} can be written as 
\begin{equation}
 e^{iP}= \prod_{j=1}^{J_2} \frac{u_j + i/2}{u_j -i/2} 
  = w^{-k}. \label{momentum2} 
\end{equation}
An expression of the energy in terms of the Bethe roots is the same as \eqref{anom}. Taking logarithms of
\eqref{bethe2} and \eqref{momentum2} and taking large $L$ expansion, we obtain
\begin{gather}
 \frac{1}{x_i} = \frac{2}{L} \sum_{j=1,j\neq i}^{J_2} 
 \frac{1}{x_i - x_j} - 2\pi (n + 2\frac{k}{M}).\quad n\in {\mathbb Z}
 \label{bethelog2}\\
 \frac{1}{L} \sum_{i=1}^{J_2} \frac{1}{x_i} = -2\pi (m +
 \frac{k}{M}),\quad m \in {\mathbb Z}
 \label{momentumlog2}
\end{gather}
where we rescale $u_i = L x_i$. Solving these equations and using \eqref{anomlargeL}, we obtain
leading terms of the anomalous dimensions in the large $L$ expansion, 
\begin{equation}
 \gamma = \frac{\lambda}{2L} (m+ \frac{k}{M})(n-m + \frac{k}{M}). 
\label{ad2}
\end{equation}
Note that $m+k/M$, $n+2k/M$ and $(n+2k/M)-(m+k/M)$ all have the same
sign.(See Appendix.) 
So, identifying $m_1 = (m+\frac{k}{M})$, $-m_2 = (n+2\frac{k}{M}) -
(m+\frac{k}{M})$ and remembering $m_2$ has the opposite sign to $m_1$, the result \eqref{ad2} agrees with
\eqref{energy2}. Also in this case, we can see that the twisted string states
correspond to the twisted sectors of the operators. 

\paragraph{$X$, $Z$ case} 
Let us look at one more example of spin chains associated with other
broken $SU(2)$s. We consider the operators consisting of $X$ and $Z$
fields, which form
a doublet of $SU(2)\subset SO(6)$ in the original ${\cal N}=4$ theory. This
$SU(2)$ is broken in the orbifold theory we are considering. The argument is
almost parallel to that of $SU(2)_L$ case. The spin chain corresponding to this $(X, Z)$ system
is a XXX spin chain of the Hamiltonian \eqref{XXX} with twisted boundary
condition, which has the following properties\footnote{From here on,
even when we change a basis of spin chain to a new set, we
denote new basis as $\uparrow$, $\downarrow$ without tilde.}.
\begin{gather}
T|\text{state}\rangle_k = w^{-k}|\text{state}\rangle_k,\\
|\uparrow_{\text{1st site}}\rangle_k = |\uparrow_{\text{(L+1)th
 site}}\rangle_k,\\
|\downarrow_{\text{1st site}}\rangle_k = w^k |\downarrow_{\text{(L+1)th
 site}}\rangle_k.
\end{gather}
The anomalous dimensions can be obtained by solving the corresponding
Bethe ansatz and the result is 
\begin{equation}
 \gamma = \frac{\lambda}{2L} (m + \frac{k}{M})(n - m). \quad m,\ n \in
  {\mathbb Z}
\end{equation}
Here, $m+k/M$, $n+k/M$ and $(n+k/M)-(m+k/M)$ have the same sign. 
Then, identifying $m_1 =m+k/M$ and $-m_3
= (n+k/M)-(m+k/M)$ and remembering that $m_3$ has the opposite sign to
$m_1$, this agrees with \eqref{energy3}. 

\section{$SU(2)$ sectors of more general ${\mathbb C}^3 /Z_M$ orbifolds.}
In this section, we show that the integrabilities of the gauge side and
the agreements between the string side and the gauge side in $SU(2)$ sectors
hold even in ${\cal N}=1$ or non-supersymmetric orbifolds.
Hence, we allow ${\mathbb Z}_M$ orbifold action arbitrary. That is, we
consider orbifold $(X,\ Y,\ Z)\sim (\omega^{p_X}X,\ \omega^{p_Y}Y,\
\omega^{p_Z}Z)$, $(p_X,p_Y,p_Z \in {\mathbb Z})$. 

\subsection{Correspondence of rational solutions}
We consider a $SU(2)$ subsector consisting of complex scalars $A$,
$B$. Let the identification by orbifolding on these fields be $(A, B) \sim (\omega^p A,
\omega^q B)$. Here, the integers $p$, $q$ take values in $[
-(M-1),(M-1)]$. In the ${\cal N}=2$ cases of the previous section,
the $SU(2)_R$ case corresponds to $A= X, \ B= Y^\dagger$ and $p=q=1$, 
the $SU(2)_L$ case to $A= X, \ B= Y$ and $p = 1,\ q = -1$
and the $(X,Z)$ case to $A=X,\ B= Z$ and $p=1,\ q=0$. 

First, let us consider the string side. We consider 2-spin circular
strings as in the case of ${\cal N}=2$. The generalization of the result
of the previous section is straightforward. The string energy in present
case becomes 
\begin{gather}
 E -L = - \frac{\lambda}{2L} m_A m_B, \nonumber\\
m_A = (\tilde{m}_A + \frac{p}{M}k), \quad 
- m_B = (\tilde{m}_B - \frac{q}{M}k), \quad 
\tilde{m}_A, \tilde{m}_B \in {\mathbb Z}.\label{genenergy}
\end{gather}
Here we assume $J_A, J_B > 0$. Hence, $m_B$ needs to have the opposite sign to
$m_A$. 

Next let us go to the field theory side. 
$A$ is $(i, \overline{i+p})$-type bifundamental
and $B$ is $(i, \overline{i+q})$-type bifundamental of the gauge groups. 
We regard $A$($B$) as up-spin(down-spin). We assume $J_A>J_B$ and take
$|\uparrow \uparrow \cdots \uparrow\rangle_k$ as ground state. 
In this theory, supersymmetry may not exist. However, vertices of
interactions are
inherited from the original theory and there are two types of
scalar potentials, one from $D$-term and the other from $F$-term. 
Like in the ${\cal N}=2$ case, the vertices from the $D$-term potential
are proportional to identity matrix in the spin chain system. The $F$-term
includes the vertices which permute neighbouring spins. These
interactions generate the phase shifts between the $L$-th site and the
first site, if $p-q\neq 0 \mod M$.(See fig.\ref{phase}.) These phase shifts can be cancelled by the
twisted boundary condition as in the previous case,
\begin{equation}
|\downarrow_{\text{1st site}}\rangle_k = 
w^{k(p-q)} |\downarrow_{\text{L+1-th site}}\rangle, 
\quad |\uparrow_{\text{1st site}}\rangle = 
|\uparrow_{\text{L+1-th site}}\rangle. \label{pqBC}
\end{equation}
The Hamiltonian of this spin chain is \eqref{XXX}. 
The cyclicity of the trace means
\begin{gather}
 \text{tr}\gamma_k \cdots A = w^{kp} \text{tr} \gamma_k A \cdots, \\
 \text{tr}\gamma_k \cdots B = w^{kq} \text{tr} \gamma_k B \cdots.
\end{gather}
So, a translation condition becomes 
\begin{equation}
 T|\text{state}\rangle_k = w^{-kp} |\text{state}\rangle_k.
\end{equation}
From these settings, we can obtain Bethe equations and a translation
condition in terms of Bethe roots, 
\begin{gather}
 \omega^{k(p-q)} \left( \frac{u_i +i/2}{u_i -i/2}\right) ^L = \prod_{j=1,j\neq i}^{J_B}
  \frac{u_i - u_j +i}{u_i -u_j -i},\\
\prod_{j=1}^{J_B}\frac{u_j +i/2}{u_j -i/2} = \omega^{-kp}.
\end{gather}
Taking logarithms of these equations and taking large $L$ limit, these
equations become
\begin{gather}
\frac{1}{x_i} =\frac{2}{L}\sum_{j=1, j\neq i}^{J_B} \frac{1}{x_i -x_j} -
 2\pi (n + k\frac{p-q}{M}), \quad n\in {\mathbb Z}\\
\frac{1}{L} \sum_{i=1}^{J_B} \frac{1}{x_i} = -2\pi (m+
 k\frac{p}{M}). \quad m\in {\mathbb Z}
\end{gather}
Here we rescale $u_i = L x_i$. The expression of anomalous dimensions in
terms of Bethe roots is the same as in the previous examples. 
Solving above equations, the anomalous dimension becomes 
\begin{equation}
 \gamma = \frac{\lambda}{2L} (m+ \frac{p}{M}k) (n - m -
  \frac{q}{M}k). \quad m,\ n \in {\mathbb Z}
\end{equation}
Notice that $(m+\frac{p}{M}k)$, $(n+ \frac{p-q}{M}k)$ and
$(n+\frac{p-q}{M}k) - (m + \frac{p}{M}k)$ have the same sign.(See Appendix.) 
Hence, identifying $m_A = m+ \frac{p}{M}k$, $-m_B = (n + \frac{p-q}{M} k)-(m
+\frac{p}{M} k)$ and remembering that $m_B$ has the opposite sign to
$m_A$, this agrees with the result of the string side \eqref{genenergy}.

\subsection{Agreement of sigma models from string side and gauge side}
In this subsection, we will see an agreement of sigma models of
both sides. From the spin chain
side, we can obtain a sigma model action by a continuum limit. On the
string side, an action which describe a string with large angular
momentum in a $S^3$ part of $S^5$ is obtained. These two actions are the
same as each other in the ${\cal N}=4$ case\cite{Kruczenski:2003gt}. We generalize this to the
orbifold cases. The argument is almost the same as that of the original
case except for boundary conditions. Hence, we will not write details
but things that are peculiar to orbifold cases.

The identification by orbifolding is $(A,\ B) \sim (\omega^p A,\ \omega^q B)$. 
The boundary condition for the spin chain is 
\begin{equation}
|\uparrow_{\text{1st site}}\rangle_k = 
w^{-k(p-q)/2} |\uparrow_{\text{L+1-th site}}\rangle_k, 
\quad |\downarrow_{\text{1st site}}\rangle_k = 
w^{k(p-q)/2} |\downarrow_{\text{L+1-th site}}\rangle_k, \label{pqBC2}
\end{equation}
where we shift an overall phase $e^{-k(p-q)/2}$ compared to
\eqref{pqBC} by using an arbitrary overall phase factor(See
\eqref{spredef}). This is done in order to make boundary condition
for $\phi(\sigma)$(see below) become well-defined. By the effect of the change of the overall phase, the
translation condition is also changed to
\begin{equation}
 T |\text{state}\rangle_k = e^{-k(p+q)/2} |\text{state}\rangle_k. \label{trans}
\end{equation}
One can introduce a continuous set of states 
\begin{equation}
 |\vec n\rangle_k = e^{i\phi/2} \cos{\theta/2} |\uparrow\rangle_k 
- e^{-i\phi/2} \sin \theta/2 |\downarrow\rangle_k.
\end{equation}
Here, $\vec n = ( \sin \theta \cos\phi, \sin\theta \sin\phi,
\cos\theta)$.
Taking large number of sites limit and labelling positions of the sites by
$\sigma$, functions $\phi(\sigma)$, $\theta(\sigma)$ become
continuous. When going around the chain, a state $|\vec n\rangle$ should
go back an initial state. Taking account of \eqref{pqBC2}, this means
that a
boundary conditions for $\phi(\sigma)$ are 
\begin{equation}
 \phi(2 \pi +\sigma) = \phi(\sigma) - 2\pi \frac{k(p-q)}{M} 
\label{phiBC}
\end{equation}
The following sigma model action is obtained by
path-integral\cite{Kruczenski:2003gt}\footnote{We take 
$0\leq \sigma \leq 2\pi$. In \cite{Kruczenski:2003gt}, $\sigma$ is taken from 0 to
$L$.}, 
\begin{equation}
 S = -\frac{L}{4\pi} \int d\sigma dt \cos\theta \dot\phi 
  -\frac{\lambda}{16 \pi L} \int d\sigma dt
((\partial_\sigma \theta)^2 + \sin^2 \theta (\partial_\sigma \phi)^2. 
\label{s-act}
\end{equation}
The momentum condition \eqref{trans} dictates that the momentum along the
$\sigma$-direction becomes
\begin{equation}
 P \equiv \frac{2\pi}{L} \int T_{01} d\sigma = -\frac{1}{2}\int \cos\theta 
\partial_{\sigma} \phi = -2\pi \frac{k(p+q)}{2M}. \label{momentum3}
\end{equation}

Next, let us move to the string side. We are interested in $R_t\times S^3$
motion of the string. $R_t$ is time direction of global coordinate of $AdS_5$ and
$S^3$ is a part of $S^5$. The metric is 
\begin{equation}
 ds^2 = -dt^2 + d\Omega_3^2 = -dt^2 +d\psi^2 +\cos\psi d\phi_1^2 + 
\sin\psi d\phi_2^2.
\end{equation}
$A,\ B$ are expressed in these coordinates as $A= \cos\psi e^{i\phi_1},\
B= \sin\psi e^{i\phi_2}$. The identification by orbifolding is $(\phi_1, \ \phi_2)\sim
(\phi_1 + 2\pi \frac{p}{M}, \ \phi_2 + 2\pi \frac{q}{M})$. 
One changes the coordinate to 
$\phi_1 = \varphi_1 + \varphi_2,\ \phi_2 =\varphi_1 -\varphi_2$, and
obtains a metric
\begin{equation}
 ds^2 = -dt^2 + d\psi^2 +d\varphi_1^2 + d\varphi_2^2 
  + 2\cos(2\psi)d\varphi_1d\varphi_2. \label{newmet}
\end{equation}
We are considering string motion with large angular momentum $L=
J_1+J_2$. In the new coordinate, this corresponds to fast motion along
$\varphi_1$. One can subtract motion of $\varphi_1$ direction from the
Polyakov action in \eqref{newmet}. A resulting sigma model action has
the same form as the action \eqref{s-act} from the spin chain, if we identify
$\varphi_2 = -\frac{1}{2}\phi,\ \psi= \frac{1}{2} \theta$ \cite{Kruczenski:2003gt}. 
Boundary conditions for $\varphi_1,\ \varphi_2$ are 
\begin{gather}
 \varphi_1(\sigma + 2\pi) = \varphi_1(\sigma) + 
 2\pi \frac{k(p+q)}{2M}, \label{phi1BC}\\
 \varphi_2(\sigma+ 2\pi) = \varphi_2(\sigma) + 
 2\pi \frac{k(p-q)}{2M}. \label{phi2BC}
\end{gather}
The Virasoro constraint imposes 
\begin{equation}
 \int d\sigma \cos(2\psi) \partial_\sigma \varphi_2 
= -\int d\sigma \partial_\sigma \varphi_1 
= - 2\pi \frac{k(p+q)}{2M}. \label{momentum4}
\end{equation}
where we use the equation \eqref{phi1BC}. Remembering $\varphi_2 =
-\frac{1}{2}\phi,\ \psi= \frac{1}{2} \theta$, the equations
\eqref{phi2BC} and \eqref{momentum4} agree with the equations
\eqref{phiBC} and \eqref{momentum3}, respectively. We confirmed the
agreement between sigma models of both sides.

\section{$SU(3)$ sectors}
In this section, we investigate $SU(3)$ subsectors of $SO(6)$. We
consider a sector of three complex scalars(coordinates) $A$, $B$ and
$C$, and take orbifold action so that identification by orbifolding
becomes $(A,\ B,\ C)\sim (e^{2\pi i\frac{p}{M}}A,\
e^{2\pi i\frac{q}{M}} B,\ e^{2\pi i\frac{r}{M}} C)$. In the gauge theory
side, we consider single trace operators of the following form, 
\begin{equation}
 \text{tr} \gamma_k A^{J_1} B^{J_2} C^{J_3} + permutations.
\end{equation}
$A$ is in $(i,\ \overline{i+p})$ bi-fundamental representation of the
gauge groups. $B$ is a $(i,\ \overline{i+q})$-type and $C$ is a $(i,\
\overline{i+r})$-type. 

The argument of the one-loop anomalous dimension matrix is almost parallel to $SU(2)$
cases. As explained in the previous section, the one-loop anomalous dimension matrix of the
untwisted sector of the operators is the same as that of ${\cal N}=4$
SYM, that is, a Hamiltonian of the
$SU(3)$ XXX spin chain, 
\begin{equation}
 D_{\text{1-loop}} = \frac{\lambda}{8\pi^2} \sum_{l=1}^{l=L} 
(1- P_{l,l+1}), \label{XXX3}
\end{equation}
where $P_{l,l+1}$ is a permutation operator between the $l$-th site and
the $l+1$-th site. A boundary condition is periodic. This model is
solvable. On the other hand, the
anomalous dimension matrices of the twisted sectors of the operators may
have the phase shifts between the $L+1$-th site and the first site. As in
the $SU(2)$ cases, these phases can be cancelled by twisted boundary
conditions. As a result, the twisted sectors are regarded as the $SU(3)$
XXX spin chains \eqref{XXX3} with the twisted boundary conditions.

\subsection{Agreement of sigma models from both sides}
Now, we extend the argument of the sigma model agreement in previous
section to the $SU(3)$ case. In the ${\cal N}=4$ case, the $SU(3)$
generalization was done in \cite{Hernandez:2004uw}. 

We regard $A$, $B$ and $C$ as spin chain states $|1\rangle$,
$|2\rangle$ and $|3\rangle$, respectively. By the same argument as the
$SU(2)$ cases, we take twisted boundary conditions,
\begin{gather}
 |1_{\text{1st-site}}\rangle_k = e^{- 2\pi i \frac{k(p-q)}{2M}} 
  |1_{\text{L+1th-site}}\rangle_k,\quad 
 |2_{\text{1st-site}}\rangle_k = e^{ 2\pi i \frac{k(p-q)}{2M}} 
|2_{\text{L+1th-site}}\rangle_k,\nonumber\\
 |3_{\text{1st-site}}\rangle_k = 
e^{ 2\pi i \frac{k}{M}(\frac{p+q}{2} -r)} 
|3_{\text{L+1th-site}}\rangle_k,
\end{gather}
A Hamiltonian of this system is \eqref{XXX3}. In this boundary condition,
the cyclicity of the trace means
\begin{equation}
 T|\text{state}\rangle_k = e^{-2\pi i \frac{k(p+q)}{2M}} 
|\text{state}\rangle_k.\label{eq73}
\end{equation}
We introduce coherent states
\begin{equation}
 |\vec n\rangle_k =\cos\frac{\theta_2}{2}\cos\frac{\theta_1}{2} 
  e^{i\phi/2}|1\rangle_k 
  - \cos\frac{\theta_2}{2}\sin\frac{\theta_1}{2} e^{-i\phi/2}|2\rangle_k 
  + \sin\frac{\theta_2}{2} e^{i\chi/2}|3\rangle_k
\end{equation}
In large number of sites limit, one obtains an action\cite{Hernandez:2004uw}
\begin{align}
 S =& -\frac{L}{4\pi} \int d\sigma dt (\sin^2\frac{\theta_2}{2} \dot\chi
 + \cos^2\frac{\theta_2}{2} \cos\theta_1 \dot\phi)\nonumber\\
 &-\frac{\lambda}{16\pi L} \int d\sigma dt ({\theta_2'}^2 
 + \cos^2 \frac{\theta_2}{2} ({\theta_1'}^2 + \sin^2 \theta_1 
 {\phi'}^2) + \sin^2\frac{\theta_2}{2} \cos^2\frac{\theta_2}{2} 
 (\chi' -\cos\theta_1 \phi')^2),\label{s-act3}
\end{align}
where the dot and prime denote the derivative of $t$ and $\sigma$,
respectively. The $SU(2)$ action of the previous section is obtain by
$\theta_2 =0$ and $\theta_1=\theta$ here. 
A periodic condition for $|\vec n\rangle$ imposes boundary conditions
for $\phi(\sigma),\ \chi(\sigma)$, 
\begin{gather}
 \phi(\sigma + 2\pi) = \phi(\sigma) - 2\pi \frac{k(p-q)}{M}, 
\label{perphi}\\
 \chi(\sigma + 2\pi) = \chi(\sigma) 
 + 2\pi \frac{k}{M}(p+q -2r). \label{perchi}
\end{gather}
The translation condition \eqref{eq73} means 
\begin{equation}
 P\equiv \frac{2\pi}{L} \int d\sigma T_{01} = -\frac{1}{2}\int d\sigma 
  (\sin^2\frac{\theta_2}{2} \chi' + \cos^2\frac{\theta_2}{2}\cos\theta_1 
  \phi') = -2\pi \frac{k(p+q)}{2M}. \label{momentum5}
\end{equation}

Next, consider the string side. In the $SU(3)$ sector, a string rotates
in $R_t\times S^5$. The relevant part of the metric is 
\begin{equation}
 ds^2 =-dt^2 + d\Omega_5^2= -dt^2 + d\theta^2 + \sin^2\theta d\phi_3^2 + \cos^2\theta 
(d\psi^2 + \cos^2 \psi d\phi_1^2 + \sin^2 \psi d\phi_2^2).
\end{equation}
$A$, $B$ and $C$ is expressed by these coordinates as 
$A = \cos\theta\cos\psi e^{i\phi_1},\ B= \cos\theta\sin\psi
e^{i\phi_2},\ C= \sin\theta e^{i\phi_3}$. 
The identification by orbifolding is expressed as $(\phi_1,\ \phi_2,\ \phi_3) \sim (\phi_1 +
2\pi \frac{p}{M},\ \phi_2 + 2\pi \frac{q}{M},\ \phi_3 + 2\pi
\frac{r}{M})$. 

Introducing next coordinates
\begin{equation}
 \phi_1 = \varphi_1 + \varphi_2,\quad \phi_2 = \varphi_1 - \varphi_2, 
  \quad \phi_3 = \varphi_1 + \alpha,
\end{equation}
then the metric becomes 
\begin{align}
 ds^2 =& -dt^2 + d\theta^2 +\sin^2\theta d\alpha^2 
  + 2\sin^2\theta d\alpha d\varphi_1 + d\varphi_1^2 \nonumber\\
 &+ \cos^2\theta (d\psi^2 + 2\cos(2\psi) d\varphi_2 d\varphi_1 
  + d\varphi_2^2).\label{newmet2}
\end{align}
In the new coordinate, the string coordinates have the following boundary
conditions, 
\begin{gather}
\varphi_1(\sigma +2\pi) = \varphi_1(\sigma) + 2\pi 
\frac{k(p+q)}{2M}, \label{phi13}\\
\varphi_2(\sigma +2\pi) = \varphi_2(\sigma) + 2\pi 
\frac{k(p-q)}{2M}, \label{phi23}\\
\alpha(\sigma +2\pi) = \alpha(\sigma) + 2\pi 
\frac{k}{M}(r - \frac{p+q}{2}). \label{alpha3}
\end{gather}
One consider a string with large angular momentum along to
$\varphi_1$ in this background. Subtracting this fast motion from the
Polyakov action in \eqref{newmet2}, the action becomes the same as that of the spin chain
\eqref{s-act3} with identifying $\theta=\frac{\theta_2}{2},\ \psi =
\frac{\theta_1}{2},\ \varphi_2 = -\frac{\phi}{2},
\ \alpha= -\frac{\chi}{2}$\cite{Hernandez:2004uw}. 

The Virasoro constraint imposes 
\begin{equation}
 \int d\sigma (\cos^2\theta 
  \cos(2\psi) \varphi_2' +\sin^2 \theta \alpha')
  = - \int d\sigma \varphi_1 ' = - \frac{2\pi}{2} 
  (m_1 + m_2 + \frac{k(p+q)}{M}),\label{vil3}
\end{equation}
where we use \eqref{phi13}. Remembering $\theta=\frac{\theta_2}{2},\
\psi = \frac{\theta_1}{2},\ \varphi_2 = -\frac{\phi}{2},\
\alpha=-\frac{\chi}{2}$, the equations \eqref{phi23}, \eqref{alpha3} and
\eqref{vil3} are equivalent to the equations \eqref{perphi},
\eqref{perchi} and \eqref{momentum5}.

\section{Conclusions}
In this paper, we have shown the agreements, at one-loop level of field
theory, between the energies of the semiclassical
strings in $AdS_5 \times S^5/{\mathbb Z}_M$ and the anomalous dimensions
of the long length operators of ${\cal N}=0,1,2$ planar orbifold field
theories originating from ${\cal N}=4$ SYM. On the
gauge theory side, we have found that the orbifold field theories have
the integrabilities in $SU(2)$ and $SU(3)$ sectors even when 
these symmetries are broken by the orbifold action. For an untwisted sector
of operators, a corresponding spin chain is a $SU(2)$($SU(3)$) XXX spin chain with periodic
boundary condition. For twisted sectors of operators, if this $SU(2)$($SU(3)$) is
global symmetry of the orbifold theory, corresponding spin chain
models are also XXX
spin chains with periodic boundary condition. On the other hand, in the
case of the broken symmetry, spin chains corresponding to twisted sectors
of operators 
become XXX spin chains with twisted boundary condition. These facts suggest
that $SU(2,2|4)$ super spin chains with
twisted boundary condition correspond to anomalous dimension matrices of 
full-sector operators in orbifold theories. This
may be the case even for the non-supersymmetric orbifold field theories. 

With the spin chains corresponding to the orbifold theories in hand, we
compared the 2-spin circular string solutions of the strings in 
$AdS_5\times S^5/{\mathbb Z}_M$ to the rational solutions of the $SU(2)$
XXX spin chain. We saw the agreement between the energies of the strings and
the anomalous dimensions of the operators at one-loop level. Especially, the strings in
the twisted sectors correspond to the `twisted sectors' of the operators. 
Furthermore, we have generalized the agreements between the sigma models of
string and gauge sides to the orbifold cases. This is done for $SU(2)$ sectors and $SU(3)$ sectors.

Finally, let us briefly discuss higher loops. 
Also at higher loops, the phase shifts born at interactions which step
$\gamma_k$ are cancelled by the same twisted boundary condition as
at one-loop level. Then, corresponding spin chain is the same as that in
${\cal N}=4$ case\cite{Serban:2004jf, Beisert:2004hm} except for twisted boundary condition. Hence, the
agreements of string energies and anomalous dimensions would hold in
higher loop level, if these hold in ${\cal N}=4$ case. It is known
that this agreement survive at 2-loop level, but there are discrepancy
at 3-loop level in ${\cal N}=4$ case\cite{Serban:2004jf,
Beisert:2004hm}. So, this would be the case for orbifold cases we consider. 

It would be interesting to expand our results to more broaden
class of operators, say full-sector of operators. 

\section*{Acknowledgements}
The author would like to thank Y.\ Imamura and T.\ Kawano for helpful comments and
reading the manuscript and Y.\ Matsuo for useful advice and
encouragement. 

\appendix

\section{Anomalous dimension matrix in untwisted sector of operators}
We consider untwisted operators consisting of $X$, $Y$ fields in ${\cal
N}=2$ case. These are
$NM\times NM$ matrix valued fields \eqref{scalar}. The
operators we consider have the form of $\text{tr} XXY\cdots$, etc. Arguments in
this section can be applied also to $SU(2)$($SU(3)$) subsectors of
Sec.~3(Sec.~4). 

\begin{figure}[htb]
 \begin{center}
  \epsfig{file=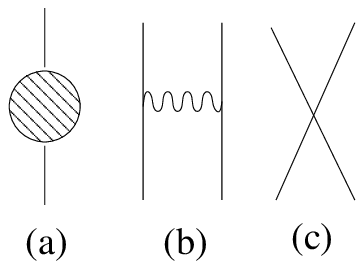}
\caption{Diagrams which contribute to $Z$-factor.}
\label{Zdia}
 \end{center}
\end{figure}

There are three types
of diagrams fig.\ref{Zdia} which contribute to renormalization $Z$-factor. 
Vertices of orbifold theory are inherited from ${\cal N}=4$ $U(NM)$ SYM. Hence, there
is one to one correspondence between diagrams of orbifold theory and
diagrams of ${\cal N}=4$ SYM. Furthermore, contribution of each diagram
to $Z$-factor in orbifold theory agrees with that in ${\cal N}=4$ $U(N)$ SYM. 
Let us explain the last statement. 

\begin{figure}[tb]
 \begin{center}
  \epsfig{file=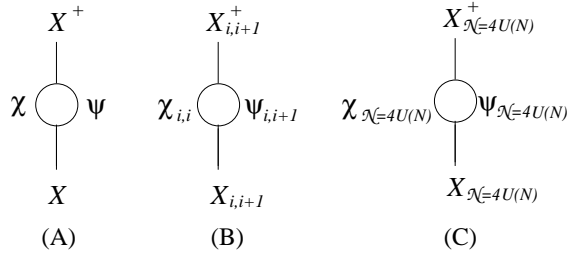}
  \caption{Diagram from yukawa coupling $\text{tr} X^\dagger\chi\psi$. A diagram
  (A) consist of $NM\times NM$ matrix valued fields \eqref{scalar} in orbifold
  theory. $\chi$ is gaugino and $\psi$ is ${\cal N}=1$ superpartner of
  $X$. This diagram (A) includes only one diagram (B) which is concerned with
  a component field $X_{i,i+1}$, which is an $N\times N$ matrix. Suffix of field
  $X_{i,i+1}$, $\chi_{i,i}$, $\psi_{i,i+1}$ represent gauge group `name'. A
  diagram (C) is a corresponding diagram of ${\cal N}=4$ $U(N)$ SYM.}
  \label{yukawa}
 \end{center}
\end{figure}

It is convenient to see diagrams in terms of component fields, which is $N\times N$
matrix valued ($X_{12},\ Y_{21},\ X_{45},$ etc). For instance, in a diagram
consisting of $NM\times NM$ matrix valued fields $X$, $\chi$ and $\psi$
of the orbifold theory (fig.\ref{yukawa}(A)), there is only one diagram
which is concerned with a field $X_{i,i+1}$ (fig.\ref{yukawa}(B)). In fig.\ref{yukawa}(B), fields which run a
loop are bifundamental fermions which are $N\times N$ matrices. 
Let us compare the diagram fig.\ref{yukawa}(B) with a diagram of ${\cal N}=4$ $U(N)$ SYM (fig.\ref{yukawa}(C)). 
In fig.\ref{yukawa}(C), fields which run a loop are adjoint fermions which
are $N\times N$ matrices and then, the value of this diagram is the same as
that of the diagram of fig.\ref{yukawa}(B). This fact is true for each
$N\times N$ valued component fields. Since the diagram
fig.\ref{yukawa}(A) is merely linear summation of the diagram
fig.\ref{yukawa}(B) for each component field, a contribution of fig.\ref{yukawa}(A) to $Z$-factor
in orbifold theory is the same as that of fig.\ref{yukawa}(C) to $Z$-factor in ${\cal
N}=4$ $U(N)$ SYM. This fact is approved for other diagrams in
fig.\ref{Zdia}. Therefore a form of anomalous dimension matrix of untwisted
operators in orbifold theory is the same as that of ${\cal N}=4$ $U(N)$ SYM. 

\section{Rational solutions of Bethe equations}
Here, we outline how to solve Bethe equations and translation condition
in large $L$ limit and obtain anomalous dimension using the
solutions. We use a method which are described in Appendix of
\cite{Lubcke:2004dg}. For details, we would like to refer to the
original paper.

The equations we consider are 
\begin{gather}
 \frac{1}{x_i} = \frac{2}{L}\sum_{j=1, j\neq i}^J \frac{1}{x_i-x_j}
  -2\pi n,\label{bethegen}\\
 \frac{1}{L} \sum_{i=1}^{J} \frac{1}{x_i} =-2\pi m,\label{momentgen}
\end{gather}
where $m$ and $n$ are not necessarily integers. 
Let us define the resolvent 
\begin{equation}
 G(x)=\frac{1}{L} \sum_{j=1}^J \frac{1}{x-x_j}.
\end{equation}
The total momentum and total energy, which is proportional to anomalous
dimension of spin chain are described in terms of the resolvent
\begin{equation}
 P = -G(0), \quad \gamma = -\frac{\lambda}{8\pi^2 L} G'(0).
\end{equation}
Multiplying \eqref{bethegen} by $\frac{1}{x-x_i}$ and summing over $i$
and using \eqref{momentgen} appropriately, 
we obtain 
\begin{equation}
 x G^2(x) = G(x) + 2\pi nxG(x) - 2\pi m,
\end{equation}
where we drop a term of subleading in $\frac{1}{L}$ expansion. This is
quadratic equation of $G(x)$ and solution is
\begin{equation}
 G(x) = \frac{1}{2x} ( 1 + 2\pi nx -\sqrt{(1+2\pi nx)^2 - 8\pi mx}). 
\end{equation}
This $G(x)$ goes to $\frac{m}{nx}$ in $x\rightarrow \infty$. 
This value should be equal to $\frac{J}{Lx}$. Hence, $\frac{J}{L} =
\frac{m}{n}$ is imposed. So, $m$ must have the same sign as
$n$. Furthermore, Bethe ansatz assume $\frac{J}{L}\leq
\frac{1}{2}$, so this imposes $|n| \geq 2|m|$. From these facts, $(n-m)$
has the same sign as $m$ and $n$. Anyway, using
the solution, we can obtain the anomalous dimension, 
\begin{equation}
 \gamma = \frac{\lambda}{2L} m(n-m).
\end{equation}

\end{document}